\documentstyle[aaspptwo,psfig]{article}

\def\etal{{\it et al. }}

\begin{document}

\title{Globular Cluster Luminosity Functions and the Hubble Constant
from WFPC2 Imaging: The Dominant Group Elliptical NGC 5846\altaffilmark{1}}

\author{Duncan A. Forbes}
\affil{Lick Observatory, University of California, Santa Cruz, CA 95064}
\and
\affil{School of Physics and Space Research, University of Birmingham,
Edgbaston, Birmingham B15 2TT, United Kingdom} 
\affil{Electronic mail: forbes@lick.ucsc.edu}

\author{Jean P. Brodie}
\affil{Lick Observatory, University of California, Santa Cruz, CA 95064}
\affil{Electronic mail: brodie@lick.ucsc.edu}

\author{John Huchra}
\affil{Harvard--Smithsonian Center for Astrophysics, 60 Garden Street,
Cambridge, Massachusetts 02138}
\affil{Electronic mail: huchra@cfa.harvard.edu}

\altaffiltext{1}{Based on observations with the NASA/ESA {\it Hubble
Space Telescope}, obtained at the Space Telescope Science Institute,
which is operated by AURA, Inc., under NASA contract NAS 5--26555}

\begin{abstract}

The {\it Hubble Space Telescope's} Wide Field and Planetary Camera 2 (WFPC2)
has several advantages over ground--based observations for the study of
globular cluster luminosity functions (GCLFs) and distance determination. Here
we present WFPC2 data on the globular clusters associated with NGC 5846. This
giant elliptical is the dominant galaxy in a small, compact group located
$\sim$ 13 Mpc beyond the Virgo cluster.  We have detected over 1200 globular
clusters in three (central, north and south) separate pointings.  The
luminosity function in each of these pointings are statistically the same,
indicating that the mean luminosity (mass) does not vary between $\sim$ 3 and
30 kpc from the galaxy center. This suggests that dynamical friction and bulge
shocking destruction processes are insignificant.  We have fit a Gaussian and
$t_5$ profile to the GCLF (of the combined pointings) and find that it is well
represented by a turnover magnitude of $m^0_V$ = 25.05 $\pm$ 0.10 and a
dispersion of $\sigma$ = 1.34 $\pm$ 0.06. Our 50\% completeness level is $\sim$
1 mag fainter than the turnover.  After applying a metallicity correction to
the `universal' GCLF turnover magnitude, we derive a distance modulus of (m--M)
= 32.32 $\pm$ 0.23. For a group velocity V$_{CMB}$ = 1883 $\pm$ 28 km s$^{-1}$,
the Hubble constant is H$_{\circ}$ = 65 $\pm$ 8 km s$^{-1}$ Mpc$^{-1}$.  

\end{abstract}


\section{Introduction}

Significant advances have been made recently in the measurement of
extragalactic distances, which are summarized in the proceedings of the {\it
Space Telescope Science Institute} workshop on `The Extragalactic Distance
Scale'. As expected, the {\it Hubble Space Telescope} (HST) has contributed to
more than one of the distance determination methods. Most of the published
results, so far, relate to the Cepheid Variable method (e.g. Freeman \etal
1994; Mould \etal 1995; Kelson \etal 1996) and the globular cluster luminosity
function (GCLF) method (e.g. Baum \etal 1995; Whitmore \etal 1995; Forbes
1996a,b).  In the case of the GCLF method, the `standard candle' is the
magnitude of the turnover, or peak, in the luminosity function of a galaxy's
globular clusters. The method can be applied to both spiral and elliptical
galaxies. The main assumption inherent in this method is that the GCLF turnover
magnitude is constant (with a value of M$_V^0$ $\sim$ --7.5). Observations with
the HST improve the precision of GCLF studies in several ways; namely, accurate
photometry to fainter magnitudes than typically obtained from the ground, no
blending of globular clusters in the central galaxy regions and very low
contamination as most stars and galaxies can be easily excluded. This improved
precision helps to address the issue of a universal GCLF and provide an
independent estimate of the Hubble constant.  

In this paper we present a GCLF of NGC 5846 from Wide Field and Planetary
Camera 2 (WFPC2) observations.  NGC 5846 is a giant elliptical galaxy at the
center of a small, compact group of galaxies. Extended X--ray emission from hot
gas has been detected out to $\sim$ 100 kpc (Biermann, Kronberg \& Schmutzler
1989). The inferred virial mass and cooling flow rate are similar to those of
M87. Harris (1991) lists a GC specific frequency (relative richness) of S$_N$ =
2.6 $\pm$ 1.9, which is somewhat low for an elliptical galaxy and much less
than that for M87 (S$_N$ = 14 $\pm$ 3).  The distance to NGC 5846 is about 30
Mpc, which places it $\sim$ 13 Mpc beyond the Virgo cluster. It lies about 30
degrees on the sky from Virgo, and out of the plane of the local supercluster.
These features make it particularly interesting for constraining galaxy bulk
flows.  In this paper we will focus on the GCLF of NGC 5846 and use it to
derive the galaxy's distance modulus. Our WFPC2 imaging consists of three
separate pointings which allows us to explore the variation of the luminosity
function with galactocentric distance.  A future paper (Forbes \etal 1997) will
explore the color and spatial properties of the NGC 5846 GC system.  

\section{Observations and Data Reduction}

We obtained F555W (V) and F814W (I) images of NGC 5846 at three separate
pointings (central, north and south) with the WFPC2 of HST. The data were
obtained in 1996 April 14 for the central pointing, April 28 for the $\sim$ 1.5
arcmin north pointing and April 24 for the $\sim$ 2.5 arcmin south pointing.
The exposure times were all 900, 1300 secs for the F555W, and 900, 1400 secs
for F814W.  After alignment, each pair of images were combined using the STSDAS
task {\it gcombine} weighted by exposure time.  

The detection of GCs was carried out using DAOPHOT (Stetson 1987) in a manner
similar to that described in Forbes \etal (1996). A signal-to-noise threshold
of 4 per pixel, `sharpness' range of 0.55 to 1.1 and roundness of --1 to 1 was
used for the initial detection of GCs. The two filters were processed
separately, so that the F555W and F814W GC candidate lists are independent.  As
we measured 2 pixel radius aperture magnitudes, an aperture correction to a
0.5$^{''}$ aperture was applied based on Table 2 of Holtzman \etal (1995), i.e.
0.39 and 0.19 for the F555W Planetary Camera (PC) and Wide Field Camera (WFC)
CCDs, and 0.54 and 0.23 for the F814W PC and WFC CCDs.  We then used the
transformations from Table 7 of Holtzman \etal (1996), for a gain setting of 7,
to convert magnitudes into the standard Johnson--Cousins V, I system.  Finally
we corrected for Galactic extinction of A$_V$ = 0.11 and A$_I$ = 0.05 (Faber
\etal 1989).  

The candidate GC lists were then compared to the known positions and counts of
hot pixels. When the photometry would be affected by more than 2\% we excluded
those GCs from the list (typically half a dozen per CCD). A FWHM size selection
was also made. In this case objects with sizes less than 1 pixel or greater
than 3 pixels were excluded.  Visually objects with FWHM size $>$ 3 pixels are
clearly galaxies. From a simulation of the large Galactic GC Omega Cen
(Grillmair 1995) we expect all GCs at the distance of NGC 5846 to be well
within the 3 pixel size limit. The typical GC in our data has a FWHM size
$\sim$ 1.6 pixels.  Finally a visual check is performed on all images to see if
any obvious spurious objects remain.  After these criteria have been applied,
we are confident that the contamination from cosmic rays, hot pixels,
foreground stars and background galaxies is small (order of a few percent) in
our object list. An example of our final selection of GCs is given in Fig. 1
(Plate ***), in which we show part of a Wide Field Camera CCD with the selected
GCs indicated.  

It is important to note that we have not made any selection based on color. In
this paper we will use only the V band data (which is of higher signal-to-noise
than the I band data).  The color--selected sample suggests a bimodal V--I
color distribution.  Examination of the luminosity function of the blue and red
subsamples reveals that there is no statistical difference between them (as
given by a K--S test).  Thus we can analyze the total V band data irrespective
of color.  

\section{Modeling}

\subsection{Completeness Function}

It is crucial for GCLF studies to quantify the ability of DAOPHOT to detect GCs
as a function of magnitude (called the completeness function).  This is
achieved by simulating GCs and then re--running DAOPHOT with the same detection
parameters and selection criteria as for the real data. As discussed below, we
find no statistical difference between the fraction of actual detected GCs in
all three pointings. This indicates that both the intrinsic GCLF {\it and} the
completeness function do not vary from one pointing to another. We have chosen
to carry out the simulation on WFC CCD 3 of the north pointing. We simulated a
total of 1000 artificial GCs with V $>$ 25.  The fraction of detected GCs
compared to those created, in 0.1 mag bins, gives the completeness function
which is shown in Fig. 2a.  

We find that all GCs brighter than V $\sim$ 25 were detected, whereas at V =
25.95 we are 50\% incomplete. The profile shape is similar to that determined
by Forbes (1996a) from short exposure time WFPC2 data of NGC 4365, and our 50\%
incompleteness level is close to that expected based with our improved
signal-to-noise data. For the subsequent analysis the completeness function is
set to zero for magnitudes fainter than V = 25.95 in order to avoid
incompleteness corrections larger than a factor of two.  

The completeness function shown in Fig. 2a is appropriate for the WFC CCDs. The
PC CCD is expected to have a lower point source sensitivity by  $\sim$0.3 mags
given its smaller pixels (Burrows \etal 1993).  We have also simulated 1000
artificial GCs in a PC CCD and derived its completeness function. It is similar
in shape to the WFC function but shifted to brighter magnitudes as expected. We
have decided to crudely correct the PC magnitude data by the ratio of the WFC
to PC completeness functions, so that the PC data resembles the WFC data.  In
this paper we will not use data from the central PC pointing as there is some
filamentary dust present but we will use the north and south PC pointings.  

\subsection{Photometric Errors}

Photometric, or measurement, errors are also important in GCLF studies. These
can cause a shift of the GCLF peak to brighter magnitudes, as some of the
fainter GCs, with relatively large errors, move into brighter magnitude bins.
This `bin jumping' effect is described in detail by Secker \& Harris (1993) and
taken into account in their maximum likelihood technique. The DAOPHOT
determined errors are fit with an exponential of the form:\\

p.e. = exp~[a~(V~--~b)]\\

\noindent
The photometric error and the fit as a function of V magnitude are shown in
Fig. 2b. There is no appreciable difference between the WFC and PC CCD
photometric errors.  Typical photometric errors are $\pm$ 0.05 mag at V = 24. 
These photometric errors (from DAOPHOT) are similar to those found by comparing
the input and measured magnitudes of GCs from simulations.  

\subsection{Background Contamination}

Ground--based GC studies are seriously affected by contamination.  In typical
$\sim$ 1$^{''}$ seeing conditions, foreground stars and background galaxies can
make up to half of any GC candidate list based solely on magnitude (e.g.
Perelmuter, Brodie \& Huchra 1995).  Our situation is significantly improved as
most stars and galaxies can be easily identified with the high spatial
resolution of WFPC2.  We typically exclude 2--4 stars per CCD, which is similar
to that expected from high Galactic latitude star counts (see Elson \& Santiago
1996) for the range 20 $<$ V $<$ 26.  In order to quantify the number of
background galaxies to V = 26 we have re--run DAOPHOT with the same detection
parameters on a 2400 sec WFPC2 image from the Medium Deep Survey (MDS; Forbes
\etal 1994). In particular the threshold is still set at 4$\sigma$ (after
determining the appropriate background noise in the MDS image), and we exclude
objects with FWHM sizes $<$ 1 pixel and $>$ 3 pixels. No color selection is
applied. We find that the number of compact galaxies is insignificant for
magnitudes V $<$ 25. Between V = 25 and 25.95 (the 50\% incompleteness level),
the number of contaminating compact galaxies in our sample approaches 20 per
WFPC2 field-of-view. The luminosity function for background unresolved objects
is shown in Fig. 3.  

\section{Results and Discussion}

As stated earlier, we will not discuss any results for GCs in the PC CCD of the
central pointing. For GCs located on the PC CCD in the other two pointings we
have corrected their luminosity function so that the PC probes to a similar
depth as the WFC CCDs. A consistency check on this correction is provided by
Fig. 4. Here we show the cumulative fraction of actual detected GCs in the
central, north and south pointings (WFC CCDs only) compared to that of the
north plus south pointing for the PC CCDs (after applying the correction
discussed in section 3.1). This figure shows that the detection fraction, as a
function of V magnitude, is similar between the PC and WFC CCDs. There are no
GCs brighter than V = 23 in the PC CCDs and at magnitudes fainter than V = 25.5
background galaxies become a strong contaminant. Including the GCs from the PC
CCDs with the WFC CCDs gives over 1200 GCs within our 50\% completeness limit.  

Forbes \etal (1996) showed that the surface density of GCs tends to flatten off
in log space near the galaxy center. The `core radius' of the GC system ranges
from 1 kpc for small ellipticals to $\sim$ 5 kpc for giant ellipticals. They
suggested that this core radius was {\it not} due to a destruction process. 
However they only had a single central pointing available to them. Here we can
critically examine the issue of GC destruction processes.  It has been
suggested that dynamical friction and bulge shocking may operate in the central
regions of galaxies (e.g. Aguliar, Hut \& Ostriker 1988; Murray \& Lin 1992). 
Dynamical friction will preferentially destroy high mass GCs, and bulge
shocking the low density (low mass) ones. Thus we might expect to see a
variation in the luminosity function (i.e. a deficit in bright or faint GCs
respectively) with galactocentric radius.  Examination of Fig. 4 shows that the
detected fraction of GCs is similar for all three pointings. This is confirmed
statistically by a K--S test which indicates that each pointing is consistent
with having been drawn from the same parent population.  This implies that the
completeness function {\it and} the intrinsic luminosity function {\it do not}
vary between the three pointings. The WFC pointings range from about 0.3 to 3
arcminutes (3--30 kpc) in galactocentric radius. Thus we confirm the suggestion
of Forbes \etal (1996) that GC destruction does not appear to be operating in
the central regions of elliptical galaxies and therefore that the present day
GC system reflects the properties of the GCs as they were formed.  

As in previous work (Forbes 1996a,b), we have used the maximum likelihood
technique of Secker \& Harris (1993) to accurately determine the GCLF peak
magnitude and dispersion.  This technique is designed to take proper account of
detection incompleteness at faint magnitudes, photometric error and background
contamination.  The fitting procedure does not use binned data but rather
treats all data points individually.  We fit both Gaussian and t$_5$ profiles
(Secker 1992).  After fitting the GCLF for the Wide Field Camera data set, we
find the best estimate and uncertainty of a Gaussian profile to be $m^0 _V$ =
25.03 $\pm$ 0.10 and $\sigma$ = 1.29 $\pm$ 0.06.  These errors represent the
collapsed one--dimensional confidence limits for one standard deviation. An
increase in the photometric errors to the upper envelope of Fig. 2b changes
$m^0 _V$ and $\sigma$ by less than 1\%.  In Fig. 5 we show the probability
contours for the Gaussian fit.  The dispersion for the $t_5$ profile is
$\sigma_t$ = 1.01, which is theoretically related to the Gaussian dispersion by
$\sigma_G \sim 1.28 \sigma_t$.  We note that NGC 5846 has a similar t$_5$
dispersion to the Milky Way GC system ($\sigma_t$ $\sim$ 1.1) and somewhat less
than is typical for Virgo ellipticals (Secker \& Harris 1993).  If the GCs on
the two PC CCDs are included in the GCLF, then the turnover magnitude is
fainter by $\sim$ 0.1 mag and the dispersion increases by 0.05 mag.  For a
$t_5$ profile, the turnover is $\sim$ 0.1 mag brighter.  The best--fit Gaussian
profile is superposed over a binned GCLF in Fig. 6. Note that the actual
fitting procedure does not use binned data.  For our field-of-view, we have
detected about 80\% of all expected GCs, i.e. $\sim$ 20\% are too faint.  Our
50\% completeness level is about 1 mag fainter than the turnover and so the
bias of the type discussed by Secker \& Harris (1992) is minimal. From their
figure 6 we have included a correction of 0.05 mag in the turnover (making it
fainter) and 0.04 in the dispersion (making it larger). The bias corrected
results are given in Table 1.  

We calculate the distance modulus from the turnover magnitude assuming the
`universal value' for the turnover is M$^0_V$ = --7.62 with an external error
of $\pm$ 0.2 (Sandage \& Tammann 1995). This universal value is then made
fainter by a metallicity--based correction following Ashman, Conti \& Zepf
(1995). In the absence of a spectroscopic metallicity for the GC system, we can
estimate the metallicity from the mean color. Combining our data from the three
pointings gives a mean V--I of $\sim$ 1.15. This corresponds roughly to [Fe/H]
= --0.29 (Couture, Harris \& Allwright 1990), and therefore a correction of
$\Delta$M$^0_V$ $\sim$ 0.35 is required. The resulting distance moduli from the
Gaussian and $t_5$ fits are also given in Table 1.  We estimate the error on
the distance modulus to be the quadrature sum of the following errors:
photometric zeropoint (0.05), charge transfer efficiency effects (0.03),
aperture correction (0.03), extinction correction (0.02), variation in M$^0_V$
(0.2) and variation in $m^0_V$ (0.10).  The error associated with the
metallicity correction of Ashman \etal (1995) is dominated by the error in
M$^0_V$.  These errors give a total of 0.23 magnitudes. Taking the average of
the fits, we find (m--M) = 32.32 $\pm$ 0.23 and a distance of 29 $\pm$ 3.3 Mpc.
NGC 5846 is the dominant elliptical in the group and 
lies at the center of the X--ray emission. It's velocity is 
within 30 km s$^{-1}$ of the mean velocity for 
35 galaxies in the group. Thus, for the corrected recession velocity of
NGC 5846, we have adopted the group velocity with respect to the cosmic
microwave background, i.e. V$_{CMB}$ = 1883 $\pm$ 28 km s$^{-1}$ (Faber \etal
1989). For a distance modulus of 32.32 $\pm$ 0.23, the resulting Hubble
constant is 65 $\pm$ 8 km s$^{-1}$ Mpc$^{-1}$. This calculation 
assumes that any correction for bulk flows affecting NGC 5846 is 
small, which is reasonable given that NGC 5846 lies $\sim$ 13 Mpc
beyond Virgo (see figure 5 of Jacoby \etal 1992). 

\section{Conclusions}

Here we present deep WFPC2 V band images of the giant elliptical NGC
5846, observed in three separate pointings. 
About 1200 globular
clusters are detected in the three pointings, with a 50\% completeness
limit of V = 25.95. This limit is about 1 magnitude beyond the
expected turnover in the globular cluster luminosity function (GCLF) and
implies we are missing only the faintest $\sim$ 20\% of globular
clusters in our field-of-view. We find that the GCLFs are
statistically the same over galactocentric radii from $\sim$ 3--30
kpc. This places constraints on the destruction of globular clusters
via dynamical friction or bulge shocking as these would be expected to
preferentially remove the high or low mass globulars respectively. 

We have fit the combined GCLF with both Gaussian and $t_5$ profiles, using the
maximum likelihood analysis of Secker \& Harris (1993). The GCLF is well
represented by a turnover magnitude of $m_V^0$ = 25.05 $\pm$ 0.1 and dispersion
$\sigma$ = 1.34 $\pm$ 0.06 (the two fitting profiles give similar results). 
This demonstrates that a turnover magnitude, accurate to 10\%, can be
determined from WFPC2 data out to $\sim$ 30 Mpc (or 13 Mpc beyond the distance
of Virgo).  The distance modulus for NGC 5846 is (m--M) = 32.32 $\pm$ 0.23,
assuming M$^0_V$ = --7.62 and after applying a metallicity correction from
Ashman \etal (1995).  Adopting the NGC 5846 group velocity, with respect to the
cosmic microwave background of 1883 km s$^{-1}$, gives a Hubble constant of 65
$\pm$ 8 km s$^{-1}$ Mpc$^{-1}$.  

\noindent
{\bf Acknowledgments}\\
We are particularly grateful to J. Secker for the use of his maximum likelihood
code and useful suggestions. We also thank C. Grillmair for helpful
discussions. This research was funded by the HST grant GO-05920.01-94A\\

\noindent
{\bf Note added in proof}\\
After finishing this paper, we became aware of the I band surface brightness
fluctuation measurements of NGC 5846 which gave (m--M) = 32.11 $\pm$ 0.10
(Blakeslee \& Tonry 1996, personal communication). For V$_{CMB}$ = 1883 km
s$^{-1}$, this corresponds to H$_{\circ}$ = 71 $\pm$ 4 km s$^{-1}$ Mpc$^{-1}$.
These values are within the errors of our measurement. \\

\newpage
\noindent{\bf References}

\noindent
Aguliar, L., Hut, P., \& Ostriker, J. P. 1988, ApJ, 335, 720\\
Ashman, K. M., Conti, A., \& Zepf, S. E. 1995, AJ, 110, 1164\\ 
Baum, W. A., \etal 1995, AJ, 110, 2537\\
Biermann, P. L., Kronberg, P. P., \& Schmutzler, T. 1989, A \& A, 208, 22\\
Burrows, C., \etal 1993, Hubble Space Telescope Wide Field and
Planetary Camera 2 Instrument Handbook, STScI\\
Couture, J., Harris, W. E., \& Allwright, J. W. B., 1990, ApJS, 73, 671\\
Faber, S. M., \etal 1989, ApJS, 69, 763\\
Forbes, D. A. 1996a, AJ, in press\\
Forbes, D. A. 1996b, AJ, in press\\
Forbes, D. A. 1997, in preparation\\
Forbes, D. A., Elson, R. A. W., Phillips, A. C., 
Illingworth, G. D. \& Koo, D. C. 1994, ApJ, 437, L17\\
Forbes, D. A., Franx, M., Illingworth, G. D., \& Carollo, C. M. 1996, 
ApJ, 467, 126\\
Freeman, W. L. \etal 1994, ApJ, 427, 628\\ 
Grillmair, C. J. 1995, personal communication\\
Harris, W. E. 1991, ARAA, 29, 543\\
Holtzman, J., \etal 1995, PASP, in press\\
Holtzman, J., \etal 1996, PASP, submitted\\
Jacoby, G. H., \etal 1992, PASP, 104, 599 (J92)\\
Kelson, D. D. \etal 1996, ApJ, 463, 26\\
Mould, J. R. \etal 1995, ApJ, 449, 413\\
Murray, S. D., \& Lin, D. N. C. 1992, ApJ, 400, 265\\
Perelmuter, J. L., Brodie, J. P., \& Huchra, J. P. 1995, AJ, 110, 620\\
Sandage, A., \& Tammann, G. A. 1995, ApJ, 446, 1\\
Secker, J. 1992, AJ, 104, 1472\\
Secker, J., \& Harris, W. E. 1993, AJ, 105, 1358 (SH93)\\
Stetson, P. B., 1987, PASP, 99, 191\\
Whitmore, B. C., Sparks, W. B., Lucas, R. A., Macchetto, F. D., \&
Biretta, J. A. 1995, ApJ, in press (W95)\\

\newpage

\begin{figure*}[p]
\caption{\label{fig1}
Grey scale WFPC2 image (see attched jpeg image) of a $\sim$ 40$^{''}$ $\times$
40$^{''}$ area located $\sim$ 2 $^{'}$ from NGC 5846. The selected globular
clusters are indicated by black squares. The galaxy is located in the upper
left direction.  
}
\end{figure*}

\begin{figure*}[p]
\centerline{\psfig{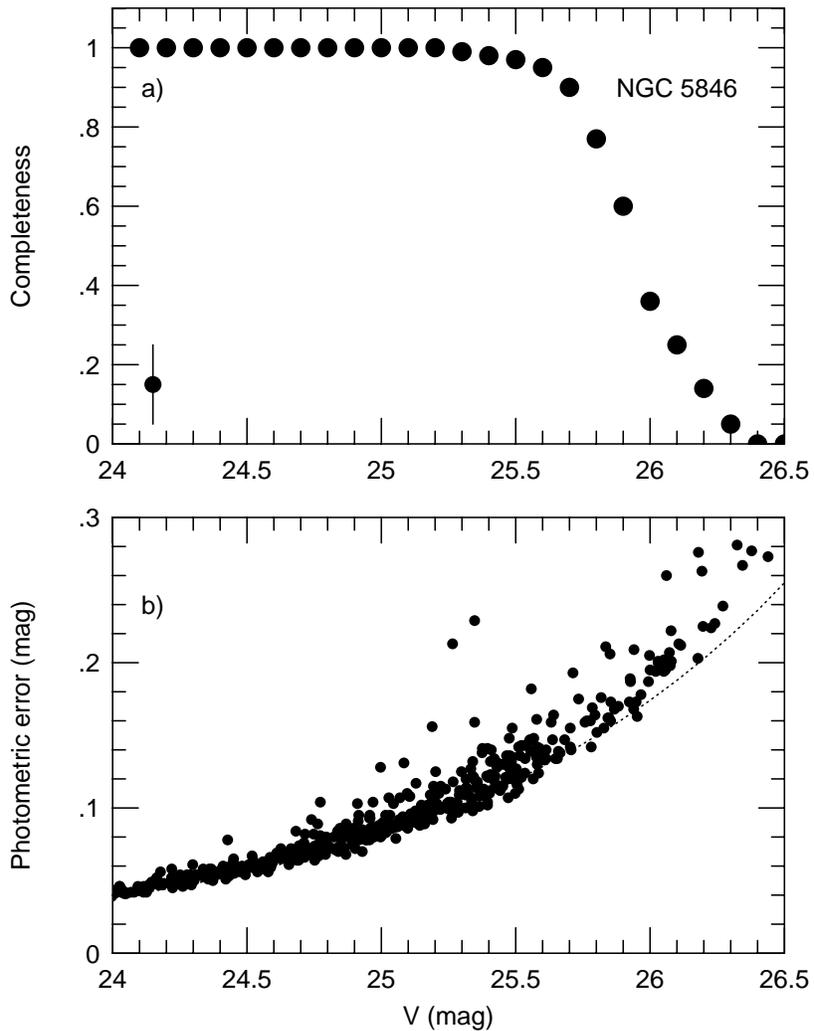}}
\caption{\label{fig2}
Completeness function for GC from simulations. Circles show the fraction of
simulated GCs detected in 0.1 magnitude bins. A typical error bar is shown in
the lower left.  Brighter than V = 25, the value is fixed to 1.0.  {\bf b)}
Photometric error as a function of GC V magnitude determined from DAOPHOT.
Circles show the data points, and the dashed line an exponential fit to the
data of the form p.e. = exp~[a~(V~--~b)].
}
\end{figure*}

\begin{figure*}[p]
\centerline{\psfig{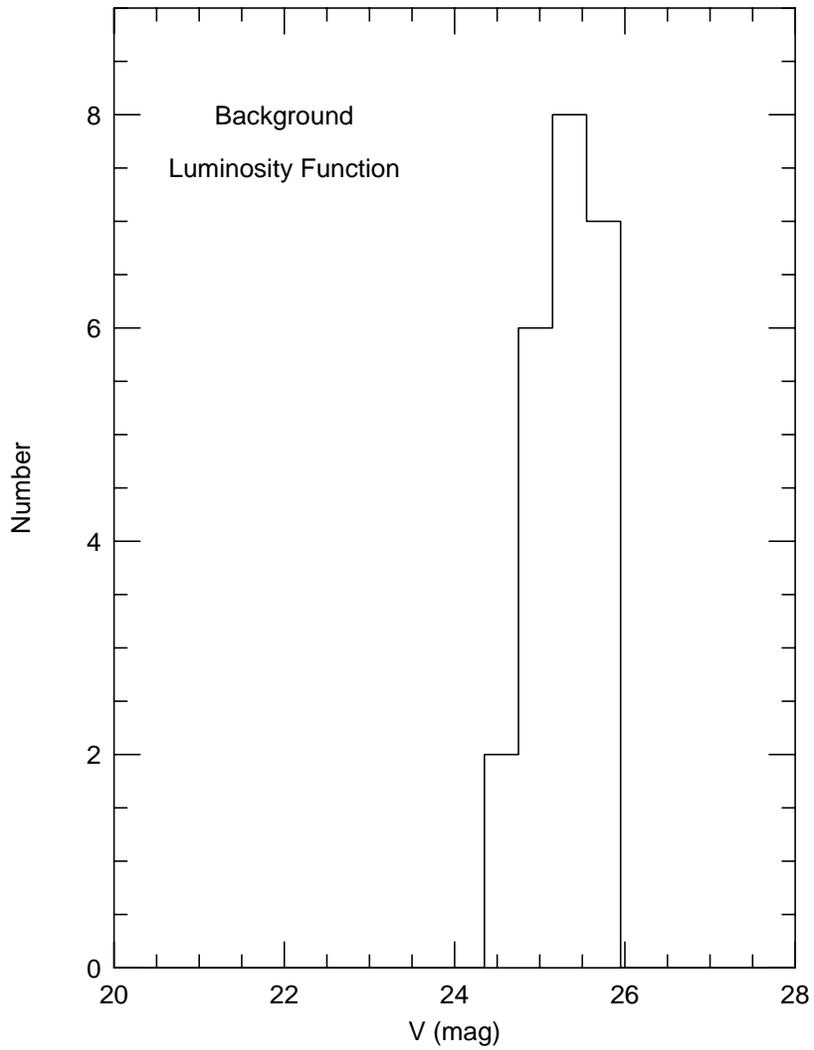}}
\caption{\label{fig3}
Luminosity function of unresolved background sources in a single high Galactic
latitude 2400 sec WFPC2 image from the Medium Deep Survey (Forbes \etal 1994).
Sources have been detected using the same parameters as the globular cluster
data.  
}
\end{figure*}

\begin{figure*}[p]
\centerline{\psfig{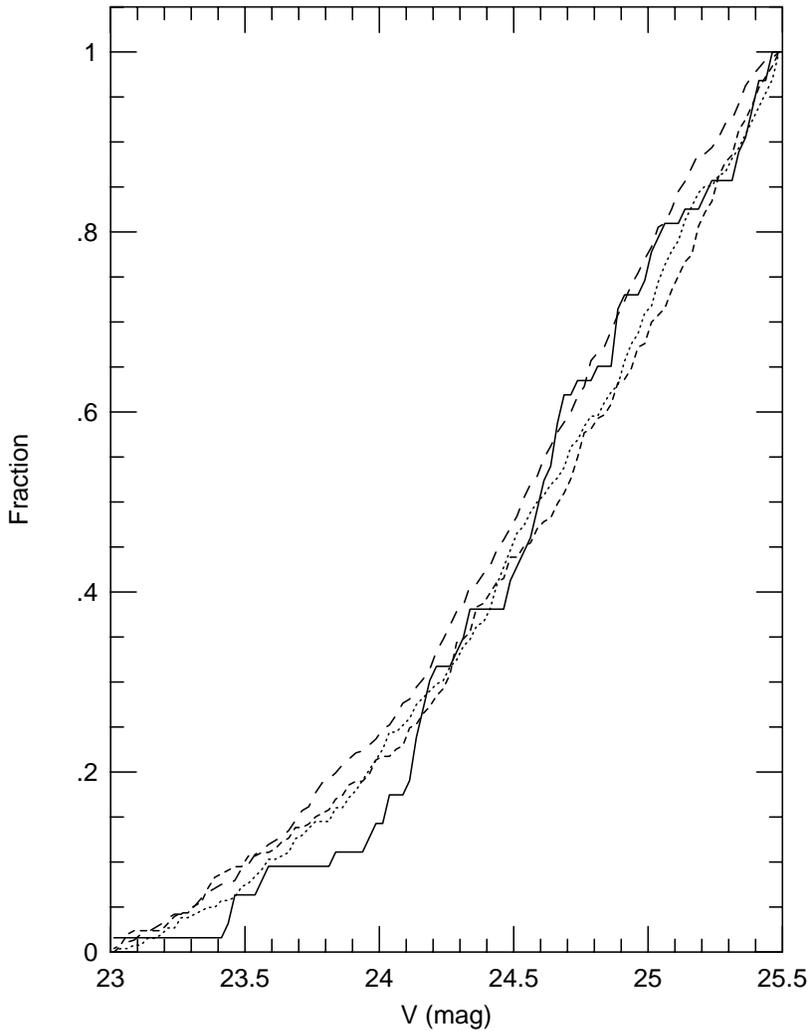}}
\caption{\label{fig4}
Detection of actual globular clusters normalized at V = 25.5.  The three
central WFC CCDs are shown by a long dash line, the WFC CCDs of the northern
pointing by a dotted line and the WFC CCDs of the southern pointing by a short
dash line.  The combined data of the north and south pointings for the PC CCDs
(after correction for their lower point source sensitivity) are shown by a
solid line.  Fainter than V = 25.5 background galaxies contribute strongly. The
fraction of actual detected globular clusters is similar over all three
pointings (a range of $\sim$ 3--30 kpc).  
}
\end{figure*}

\begin{figure*}[p]
\centerline{\psfig{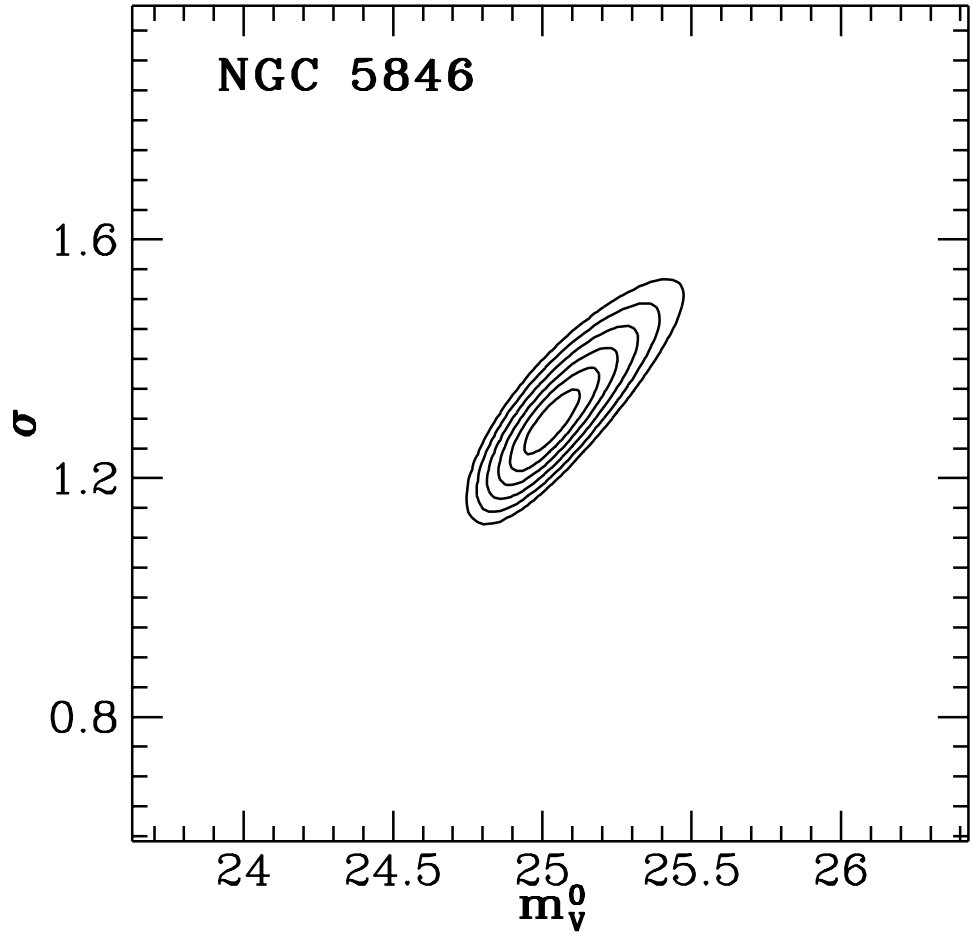}}
\caption{\label{fig5}
Probability contours for the turnover magnitude and dispersion for a Gaussian
fit from the maximum likelihood code of Secker \& Harris (1993). Contours
represent 0.5 to 3 standard deviations probability limits from the best
estimate (see Table 1).
}
\end{figure*}

\begin{figure*}[p]
\centerline{\psfig{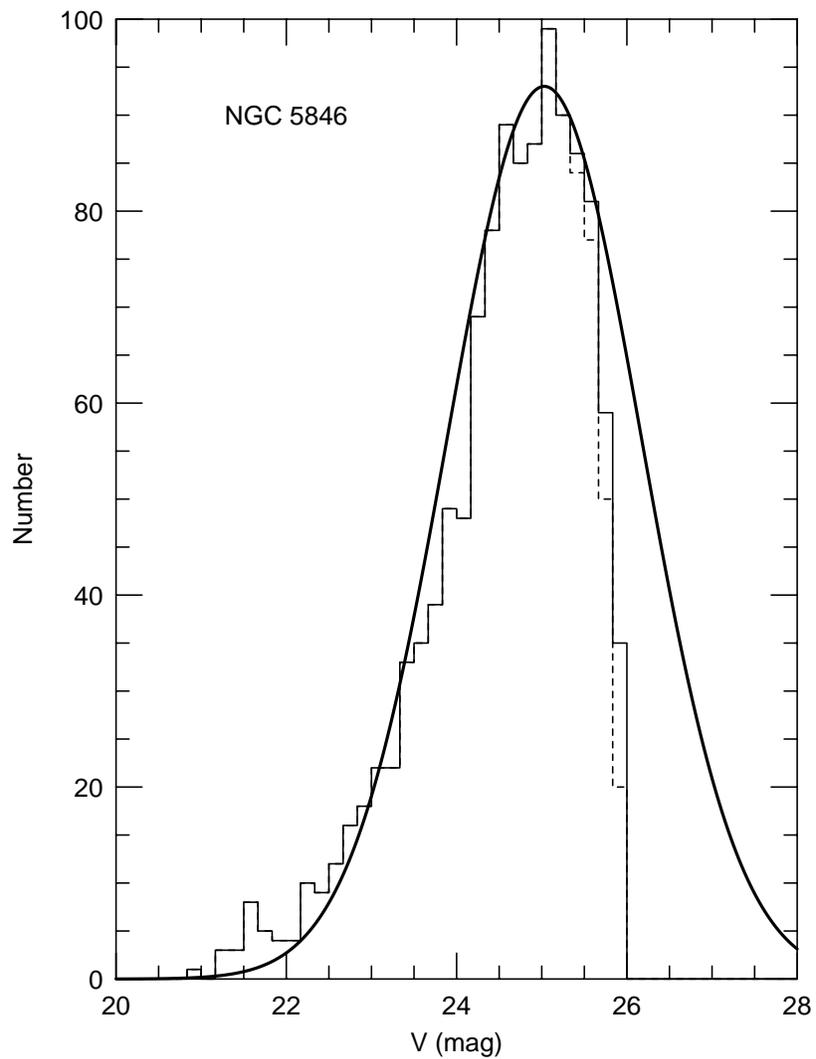}}
\caption{\label{fig6}
Globular cluster luminosity function for NGC 5846. The raw data are shown by a
dashed line, and by a thin solid line after a completeness correction has been
applied. The maximum likelihood best--fit Gaussian profile, which includes the
effects of photometric error and background contamination, is superposed as a
thick solid line.  Note that the fitting procedure does not use binned data.  
}
\end{figure*}

\clearpage
\begin{figure*}
\centerline{\psfig{figure=table1.epsi,width=300pt}}
\end{figure*}

\end{document}